\documentclass[sigconf]{acmart}
\AtBeginDocument{%
  \providecommand\BibTeX{{%
    \normalfont B\kern-0.5em{\scshape i\kern-0.25em b}\kern-0.8em\TeX}}}

\copyrightyear{2021}
\acmYear{2021}
\setcopyright{acmlicensed}\acmConference[CIKM '21]{Proceedings of the 30th ACM International Conference on Information and Knowledge Management}{November 1--5, 2021}{Virtual Event, QLD, Australia}
\acmBooktitle{Proceedings of the 30th ACM International Conference on Information and Knowledge Management (CIKM '21), November 1--5, 2021, Virtual Event, QLD, Australia}
\acmPrice{15.00}
\acmDOI{10.1145/3459637.3482286}
\acmISBN{978-1-4503-8446-9/21/11}

\usepackage{booktabs}
\usepackage{threeparttable}
\usepackage{float}
\usepackage{multirow}
\usepackage{textcomp}
\usepackage{caption}
\usepackage{booktabs}
\usepackage{bm}
\usepackage{makecell}
\usepackage{url}
\usepackage[labelformat=simple]{subcaption}
\usepackage{CJKutf8}

\usepackage{enumitem}

\newcommand{\ie}{\textit{i.e.}}
\newcommand{\eg}{\textit{e.g.}}

\newcommand{\minitab}[2][l]{\begin{tabular}{#1}#2\end{tabular}}
\newcommand\blfootnote[1]{%
\begingroup
\renewcommand\thefootnote{}\footnote{#1}%
\addtocounter{footnote}{-1}%
\endgroup
}

\begin{document}
\fancyhead{}

\title{Pre-training for Ad-hoc Retrieval: Hyperlink is Also You Need}

\author{Zhengyi Ma$^{1\dagger}$, Zhicheng Dou$^{1}$, Wei Xu$^{1}$, Xinyu Zhang$^{2}$, Hao Jiang$^{2}$, Zhao Cao$^{2}$, Ji-Rong Wen$^{1,3,4}$}

\affiliation{$^1$Gaoling School of Artificial Intelligence, Renmin University of China, Beijing, China}

\affiliation{$^2$Distributed and Parallel Software Lab, Huawei}

\affiliation{$^3$Beijing Key Laboratory of Big Data Management and Analysis Methods, Beijing, China}

\affiliation{$^4$Key Laboratory of Data Engineering and Knowledge Engineering, MOE, Beijing, China}

\email{{zymaa,dou}@ruc.edu.cn}

\renewcommand{\shortauthors}{Ma and Dou, et al.}
\renewcommand{\authors}{Zhengyi Ma, Zhicheng Dou, Wei Xu, Xinyu Zhang, Hao Jiang, Zhao Cao, and Ji-Rong Wen}

\begin{abstract}
Designing pre-training objectives that more closely resemble the downstream tasks for pre-trained language models can lead to better performance at the fine-tuning stage, especially in the ad-hoc retrieval area. 
Existing pre-training approaches tailored for IR tried to incorporate weak supervised signals, such as query-likelihood based sampling, to construct pseudo query-document pairs from the raw textual corpus. 
However, these signals rely heavily on the sampling method. For example, the query likelihood model may lead to much noise in the constructed pre-training data. \blfootnote{$\dagger$ This work was done during an internship at Huawei.}
In this paper, we propose to leverage the large-scale hyperlinks and anchor texts to pre-train the language model for ad-hoc retrieval.
Since the anchor texts are created by webmasters and can usually summarize the target document, it can help to build more accurate and reliable pre-training samples than a specific algorithm.
Considering different views of the downstream ad-hoc retrieval, we devise four pre-training tasks based on the hyperlinks.
We then pre-train the Transformer model to predict the pair-wise preference, jointly with the Masked Language Model objective.
Experimental results on two large-scale ad-hoc retrieval datasets show the significant improvement of our model compared with the existing methods.

\end{abstract}



\maketitle

\section{Introduction}  \label{section:Introduction}

\begin{figure}
\centering
\includegraphics[width=\linewidth]{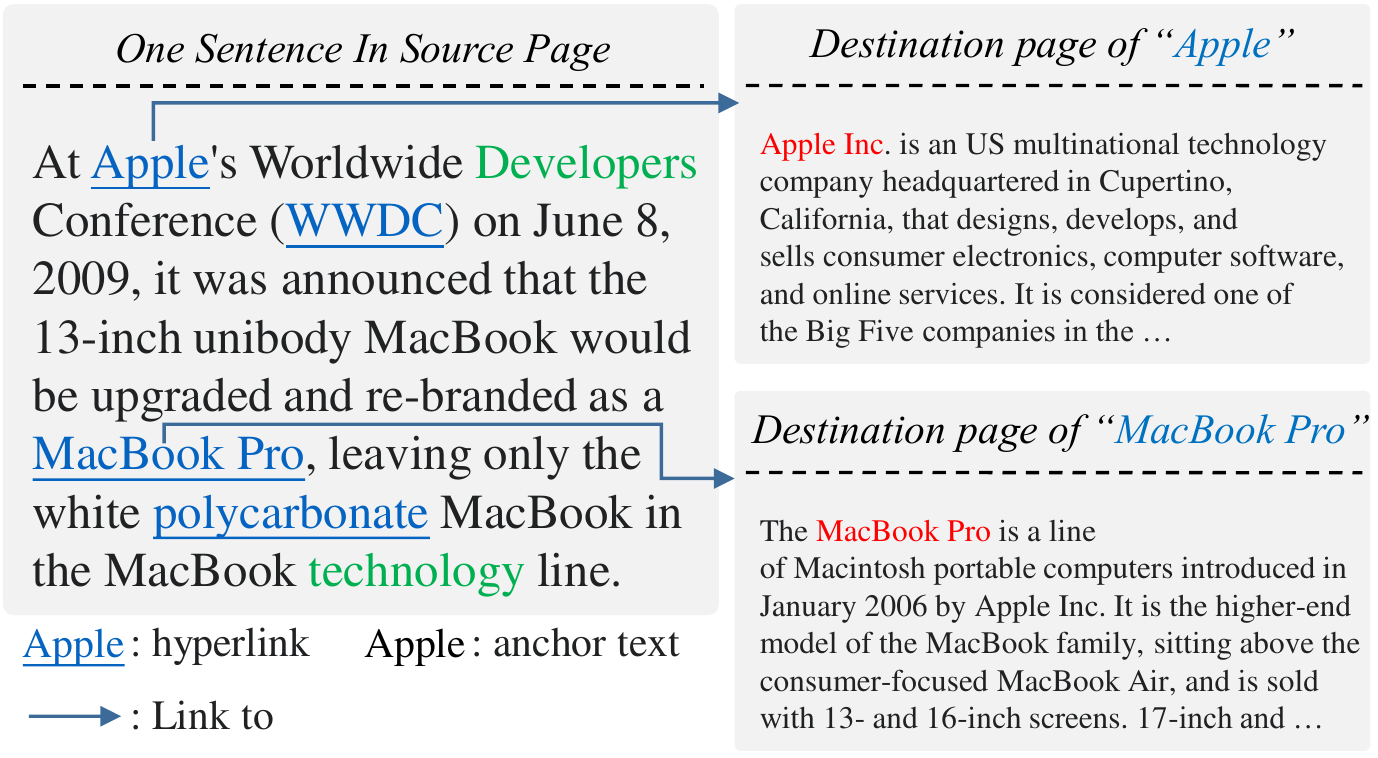}
\caption{An example of the anchor-document relations approximate relevance matches between query-document. }
\label{fig:intro_example}
\end{figure}

Recent years have witnessed the great success of many pre-trained language representation models in the natural language processing~(NLP) field~\cite{DBLP:conf/naacl/DevlinCLT19,gpt,gpt-2,DBLP:conf/iclr/ClarkLLM20}. 
Pre-trained on large-scale unlabeled text corpus and fine-tuned on limited supervised data, these pre-trained models have achieved state-of-the-art performances on many downstream NLP tasks~\cite{DBLP:conf/nips/SutskeverVL14,DBLP:conf/emnlp/SocherPWCMNP13,DBLP:conf/conll/SangM03}. 
The success of pre-trained models has also attracted more and more attention in IR community~\cite{DBLP:journals/corr/abs-1910-14424,DBLP:conf/iclr/ChangYCYK20,10.1145/3437963.3441777,DBLP:journals/corr/abs-1903-10972,DBLP:journals/corr/abs-2007-00808}. 
For example, many researchers have begun to explore the use of pre-trained language models for the ad-hoc retrieval task, which is one of the most fundamental tasks in IR.
The task aims to return the most relevant documents given one query solely based on the query-document relevance. 
Studies have shown that leveraging the existing pre-trained models for fine-tuning the ranking model over the limited relevance judgment data is able to achieve better retrieval effectiveness~\cite{DBLP:journals/corr/abs-1910-14424,DBLP:journals/corr/abs-1901-04085,DBLP:conf/ecir/GaoDC21,DBLP:journals/corr/abs-1903-10972}.

Although existing methods of fine-tuning ranking models over pre-trained language models have been shown effective, the pre-training objectives tailored for IR are far from being well explored.
Recently, there have been some preliminary studies on this direction~\cite{DBLP:conf/iclr/ChangYCYK20,10.1145/3437963.3441777}. 
For example, \citet{10.1145/3437963.3441777} proposed to sample word sets from documents as pseudo queries based on the query likelihood, and use these word sets to simulate query document relevance matching. 
Different from existing studies, in this work, we propose to \textbf{leverage the correlations and supervised signals brought by hyperlinks and anchor texts, and design four novel pre-training objectives to learn the correlations of query and documents for ad-hoc retrieval}. 

Hyperlinks are essential for web documents to help users navigating from one page to another. Humans usually select some reasonable and representative terms as the anchor text to describe and summarize the destination page. We propose to leverage hyperlinks and anchor texts for IR-oriented pre-training, because: 
(1) Since anchor texts are usually short and descriptive, based on the classical anchor intuition, \textbf{anchor texts share similar characteristics with web queries, and the anchor-document relations approximate relevance matches between query and documents}~\cite{DBLP:conf/www/0001PL15,DBLP:conf/sigir/DouSNW09,10.1145/3366423.3380131,DBLP:conf/ecir/DaiD10,DBLP:conf/sigir/YiA10}. For example, as shown in Figure~\ref{fig:intro_example}, the anchor text ``MacBook Pro'' is a reasonable query for the introductory page of itself. 
(2) Anchor texts are created and filtered by web masters~(\ie, humans), rather than generated by a specific model automatically. Thus, they can provide more accurate and reliable summarized information of one page, which further brings stronger supervised signals for pre-training. Besides, it can reflect user's information need, and help to model the matching between user needs and documents.
(3) Anchor texts can bring terms that are not in the destination page, while the existing methods mostly use the document terms for describing the document. In this way, the model can use more abundant information for capturing semantics and measuring relevance.
(4) Hyperlinks widely exist on web pages and are cost-efficient to collect, which can provide large-scale training data for pre-training models. 
In summary, hyperlinks are appropriate for pre-training tailored for IR, and easy to obtain. 

However, straightly building anchor-document pairs to simulate query-document relevance matching may hurt the accuracy of neural retrieval models, since there exist noises even spams in hyperlinks~\cite{10.1145/3366423.3380131,DBLP:conf/sigir/DehghaniZSKC17}. Besides, the semantics of short anchor texts could be insufficient. For example, as shown in Figure~\ref{fig:intro_example}, the single term of ``Apple'' is not a suitable query for the page of ``apple company'', since ``Apple'' can also refer to pages about ``apple fruit''. However, by considering the whole sentence containing the anchor ``Apple'', we could build more informative queries to describe the page, such as ``Apple technology''. This indicates that we should leverage the context semantics around the anchor texts for building more accurate anchor-based pre-training data.

Based on the above observation, we propose a pre-training framework \textbf{HARP}, which focuses on designing \textbf{P}re-training objectives for ad-hoc \textbf{R}etreival with \textbf{A}nchor texts and \textbf{H}yperlinks. 
Inspired by the self-attentive retrieval architecture~\cite{DBLP:journals/corr/abs-1901-04085}, we propose to firstly pre-train the language representation model with supervised signals brought by hyperlinks and anchor texts, and then fine-tune the model parameters according to downstream ad-hoc retrieval tasks. 
The major novelty lies in the pre-training stage. 
In particular, we carefully devise four self-supervised pre-training objectives for capturing the anchor-document relevance in different views: representative query prediction, query disambiguation, representative document prediction, and anchor co-occurrence modeling. 
Based on the four tasks, we can build a large number of pair-wise query-document pairs based on hyperlinks and anchor texts.
Then, we pre-train the Transformer model to predict pairwise preference jointly with Masked Language Model~(MLM) objective. 
Via such a pre-trained method, HARP can effectively fuse the anchor-document relevance signal data, and learn context-aware language representations. Besides, HARP is able to characterize different situations of ad-hoc retrieval during the pre-training process in a general way.
Finally, we fine-tune the learned Transformer model on downstream ad-hoc retrieval tasks to evaluate the performance.

We pre-train the HARP model on English Wikipedia, which contains tens of millions of well-formed wiki articles and hyperlinks. At the fine-tuning stage, we use a ranking model with the same architecture as the pre-trained model. We use the parameters of the pre-trained model to initialize the ranking model, and fine-tune the ranking model on two open-accessed ad-hoc retrieval datasets, including MS-MARCO Document Ranking and Trec DL. Experimental results show that HARP achieves state-of-the-art performance compared to a number of competitive methods.

Our contributions are three-fold: 
(1) We introduce the hyperlinks and anchor texts into pre-training for ad-hoc retrieval. By this means, our method can leverage the supervised signal brought by anchor-document relevance, which is more accurate and reliable than the existing methods based on specific sampling algorithms.
(2) We design four self-supervised pre-training objectives, including representative query prediction, query disambiguation modeling, representative document prediction, and anchor co-occurrence modeling to pre-train the Transformer model. In such a way, we are able to simulate the query-document matching at the pre-training stage, and capture the relevance matches in different views.  
(3) We leverage the context semantics around the anchors instead of using the anchor-document relevance straightly. This helps to build more accurate pseudo queries, and further enhance the relevance estimation of the pre-trained model.

\section{Related Work}\label{sec:related_work}

\subsection{Pre-trained Language Models}

In recent years, pre-trained language models with deep neural networks have dominated across a wide range of NLP tasks~\cite{DBLP:conf/naacl/PetersNIGCLZ18,DBLP:conf/naacl/DevlinCLT19,DBLP:conf/nips/YangDYCSL19,DBLP:conf/aaai/ZhuZNLD21}. They are firstly pre-trained on a large-scale unlabeled corpus, and fine-tuned on downstream tasks with limited data. With the strong ability to aggregate context, Transformer~\cite{DBLP:conf/nips/VaswaniSPUJGKP17} becomes the mainstream module of these pre-trained models. 
Some researchers firstly tried to design generative pre-training language models based on uni-directional Transformer~\cite{gpt,DBLP:conf/nips/YangDYCSL19,gpt-2}. 
To model the bi-directional context, \citet{DBLP:conf/naacl/DevlinCLT19} pre-trained BERT, which is a large-scale bi-directional Transformer encoder to obtain contextual language representations. Following BERT, many pre-trained methods have achieved encouraging performance, such as robust optimization~\cite{DBLP:journals/corr/abs-1907-11692}, parameter reduction~\cite{DBLP:conf/iclr/LanCGGSS20}, discriminative training~\cite{DBLP:conf/iclr/ClarkLLM20}, and knowledge incorporation~\cite{DBLP:conf/coling/SunSQGHHZ20,DBLP:conf/acl/ZhangHLJSL19}. Inspired by the powerful capacity of BERT for modeling language representations, the IR community has also explored to apply pre-trained models for better measuring the information relevance. By concatenating the query and document with special tokens and feeding them into BERT, many methods has achieved great performance by fine-tuning with BERT~\cite{DBLP:journals/corr/abs-1910-14424,DBLP:journals/corr/abs-1901-04085,qiao2019understanding,DBLP:conf/sigir/DaiC19,DBLP:journals/corr/abs-1903-10972,DBLP:conf/sigir/WeiHZ20,DBLP:conf/ecir/GaoDC21,DBLP:conf/sigir/SuDZQW21}.

\subsection{Pre-training Objectives for IR}
Although fine-tuning the downstream IR tasks with pre-trained models has achieved promising results, designing a suitable pre-training objective for ad-hoc retrieval has not been well explored. There have been several successful pre-training tasks for NLP, such as masked language modeling~\cite{Taylor1953ClozePA,DBLP:conf/naacl/DevlinCLT19}, next sentence prediction~\cite{DBLP:conf/naacl/DevlinCLT19}, permutation language modeling~\cite{DBLP:conf/nips/YangDYCSL19} and replaced token detection~\cite{DBLP:conf/iclr/ClarkLLM20}. However, they are designed to model the general contextual dependency or sentence coherence, not the relevance between query-document pairs. A good pre-training task should be relevant to the downstream task for better fine-tuning performance~\cite{DBLP:conf/iclr/ChangYCYK20}. 
Some researchers proposed to pre-train on a large-scale corpus with Inverse Cloze Task (ICT) for passage retrieval, where a passage is treated as the document and its inner sentences are treated as queries~\cite{DBLP:conf/acl/LeeCT19,DBLP:conf/iclr/ChangYCYK20}. \citet{DBLP:conf/iclr/ChangYCYK20} also designed Body First Selection and Wiki Link Prediction to capture the inner-page and inter-page semantic relation. \citet{10.1145/3437963.3441777} proposed Representative Words Prediction (ROP) task for pre-training in a pair-wise way. They assumed that the sampled word set with higher query likelihood is a more ``representative'' query. Then, they train the Transformer encoder to predict pairwise scores between two sampled word sets, and achieve state-of-the-art performance.

Different from the above approaches, we propose using the correlations brought by hyperlinks and anchor texts as the supervised signals for the pre-trained language model. 
Hyperlinks and anchor texts have been used in various existing IR studies, including ad-hoc retrieval~\cite{10.1145/3366423.3380131}, query refinement~\cite{DBLP:conf/www/KraftZ04}, document expansion~\cite{DBLP:conf/sigir/DouSNW09}, and query suggestion~\cite{DBLP:conf/wsdm/DangC10}. However, none of them consider using hyperlinks to design the pre-training objectives for IR.
Since hyperlinks widely exist in web documents and can bring complementary descriptions of the target documents, we believe they can bring stronger and more reliable supervised signals for pre-training, which further improve downstream ad-hoc retrieval performance. 

\begin{figure}
\centering
\includegraphics[width=\linewidth]{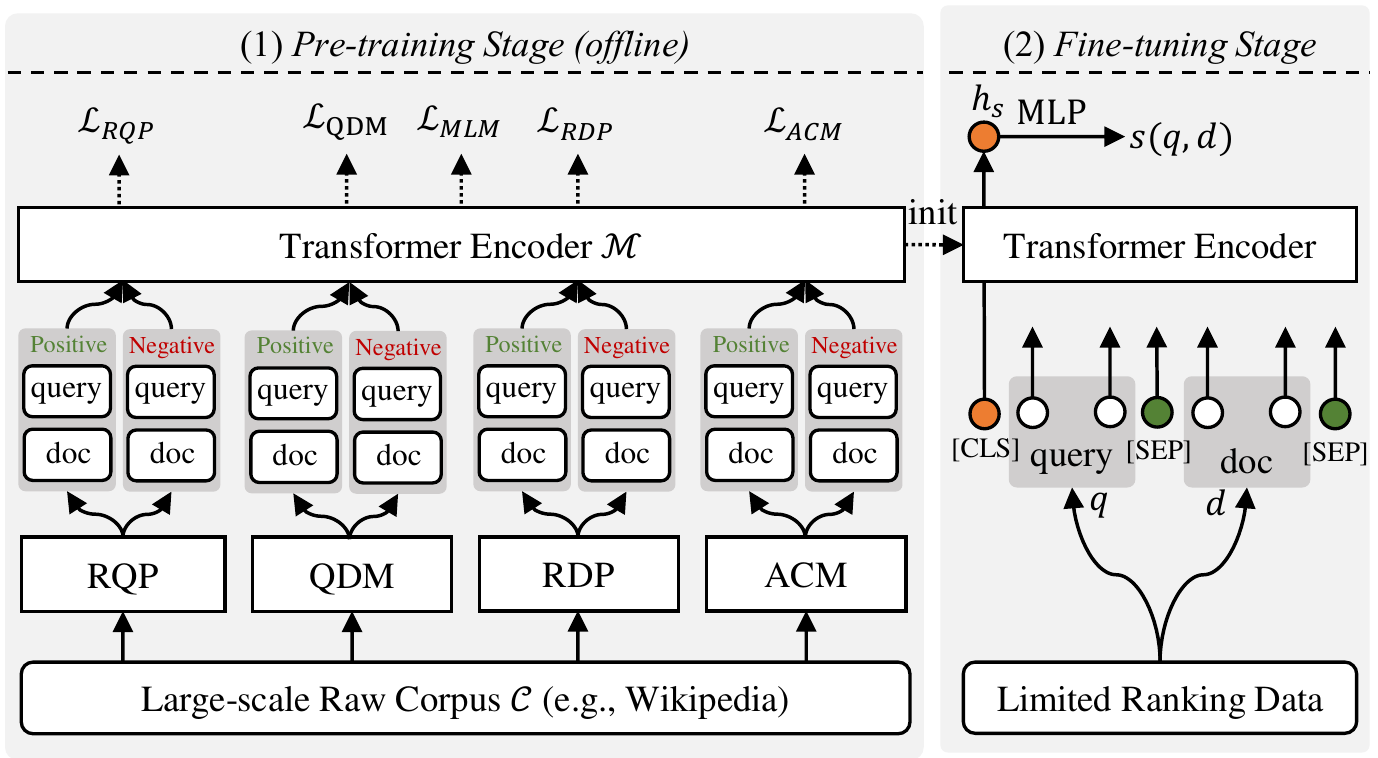}
\caption{The two-stage architectures of HARP, which consists of (1) pre-training stage, and (2) fine-tuning stage.   }
\label{fig:two_stage_framework}
\end{figure}

\begin{figure*}
\centering
    \includegraphics[width=0.9\textwidth]{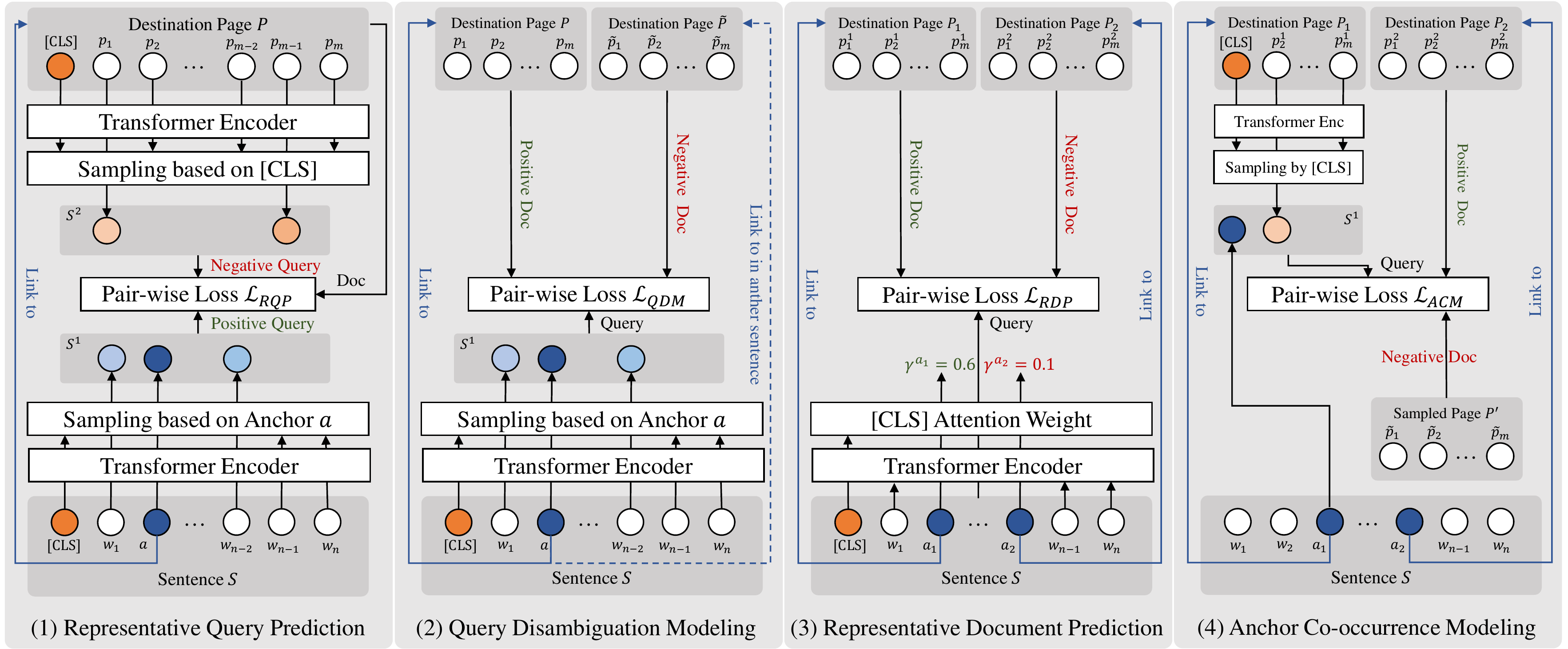}
    \caption{The proposed four pre-training tasks based on hyperlinks: (1) Representative Query Prediction, (2) Query Disambiguation Modeling, (3) Representative Document Prediction, and (4) Anchor Co-occurrence Modeling.}
\label{figure_model}
\end{figure*}

\section{Methodology}\label{sec:model}

The key idea of our approach is to leverage the hyperlinks and anchor texts for designing better pre-training objectives tailored for ad-hoc retrieval, and further improve the ranking quality of the pre-trained language model. 
To achieve this, we design a framework HARP.
As shown in Figure~\ref{fig:two_stage_framework}, the framework of HARP can be divided into two stages: (1) \textit{pre-training} stage and (2) \textit{fine-tuning} stage.
In the first stage, we design four pre-training tasks to build the pseudo query-document pairs from the raw corpus with hyperlinks, then pre-train the Transformer model with the four pre-training objectives jointly with the MLM objective.
In the second stage, we use the pre-trained model of the first stage to initialize the ranking model, then fine-tune it on the limited retrieval data for proving the effectiveness of our pre-trained model.


In this section, we first provide an overview of our proposed model HARP in Section~\ref{sec:overview}, consisting of two stages of pre-training and fine-tuning. Then, we will give the details of the pre-training stage in Section~\ref{sec:pre-train}, and the fine-tuning stage in Section~\ref{sec:fine-tune}.

\subsection{The Overview of HARP}\label{sec:overview}

We briefly introduce the two-stage framework of our proposed HARP as follows.

\subsubsection{Pre-training Stage}

As shown in Figure~\ref{fig:two_stage_framework}, in the pre-training stage, we pre-train the Transformer model to learn the query-document relevance based on the hyperlinks and anchor texts. Thus, the input of this stage is the large-scale raw corpus $\mathcal{C}$ containing hyperlinks, and the output is the pre-trained Transformer model $\mathcal{M}$. To achieve this, we design four pre-training tasks to capture different views of the anchor-document relevance, generate pseudo query-document pairs to simulate the downstream ad-hoc retrieval task, and pre-train the Transformer model toward these four objectives jointly with the MLM objective. After the offline pre-training on the raw corpus, the Transformer model can learn the query-document matching from the pseudo query-document pairs based on hyperlinks. Thus, it can achieve better performance when applied to the fine-tuning stage. 

Based on the above assumptions, we formulate the pre-training stage as follows: Suppose that in a large corpus $\mathcal{C}$ (\eg, Wikipedia), we can obtain many textual sentences. We denote one sentence as $S = (w_1,w_2,\cdots,w_n)$, where $w_i$ is the $i$-th word in $S$. In sentence $S$, some words are anchor texts within hyperlinks. We use $A = ((a_1,P_1), (a_2,P_2),\cdots, (a_m,P_m))$ to denote the set of anchor texts in sentence $S$ , where $a_i$ denotes the $i$-th anchor word in the sentence, and $P_i$ is the destination page. Figure~\ref{fig:intro_example} shows an example of anchors in one sentence. For notation simplicity, we treat the multi-words anchor texts as one phrase in the word sequence. In our pre-training corpus with anchors, one source sentence can link to one or more different destination pages using different anchor texts. In the meanwhile, a destination page can also be linked by several source sentences using different anchor texts. In fact, these characteristics of hyperlinks are leveraged in our designed pre-trained tasks to simulate some situations of ad-hoc retrieval. Based on the dataset $\mathcal{C}$, we train a Transformer model $\mathcal{M}$ on this corpus based on four pre-training tasks. The output of the pre-training stage is the model $\mathcal{M}$. Since the pre-training does not depend on any ranking data, the pre-training stage can be done offline for obtaining a good language model $\mathcal{M}$ from the large-scale corpus.

\subsubsection{Fine-tuning Stage}

As shown in Figure~\ref{fig:two_stage_framework}, in the fine-tuning stage, we use the pre-trained Transformer model to measure the relevance between a query and a document. Fine-tuned on the limited ranking data, our model can learn the data distribution of the specific downstream task and be used for ranking. The formulation of the fine-tuning stage is the same as the ad-hoc retrieval task. Given a query $q$ and a candidate document $d$, we learn a score function $s(q,d)$ to measure the relevance between $q$ and $d$. Then, for each candidate document $d$, we calculate the relevance score of them and return the documents with the highest scores. Specifically, we concatenate the query and document together, and feed them into the Transformer model. Note that the parameter and embeddings of this Transformer model are initialized by the pre-trained model $\mathcal{M}$ in the first stage. Then, we calculate the representations of the [CLS] token at the sequence head and apply a multi-layer perception(MLP) function over this representation to generate the relevance score.

\subsection{Pre-training based on Hyperlinks}\label{sec:pre-train}

In the pre-training stage, a pre-training task that more closely resembles the downstream task can better improve the fine-tuning performance. As we introduced in Section~\ref{section:Introduction}, the relations between anchor texts and documents can match the relevance of query and documents. Thus, we can leverage these supervised signals brought by anchor texts to build reliable pre-training query-document pairs. Training on these pairs, the model can learn the query-document matching in the pre-training stage, further enhance the downstream retrieval tasks. To achieve this, we design four pre-training tasks based on hyperlinks to construct different loss functions. The architecture of the four pre-training tasks is shown in Figure~\ref{figure_model}. These four tasks try to learn the correlation of ad-hoc retrieval in different views. Thus, the focus of each task is how to build the query-document pair for pre-training. 
In the following, we will present the proposed four pre-training tasks in detail.

\subsubsection{Representative Query Prediction~(RQP)}\label{sec:RQP}
Based on the classic anchor intuition, the relation between anchor texts and the destination page can approximate the query-document relevance~\cite{DBLP:conf/www/0001PL15,DBLP:conf/sigir/DouSNW09,10.1145/3366423.3380131,DBLP:conf/ecir/DaiD10,DBLP:conf/sigir/YiA10}. Therefore, our first idea is that the anchor texts could be viewed as a more representative query compared to the word set $S^2$ directly sampled from the destination page.
However, since the anchor texts are usually too short, the semantics they carry could be limited~\cite{DBLP:conf/sigir/ZhouDW20,DBLP:conf/sigir/LevelingJ10,DBLP:conf/wsdm/WuWZCDS14}.
Fortunately, with the contextual information in the anchor's corresponding sentence $S$, we can build a more informative pseudo query $S^1$ with not only the anchor text, but also the contexts in the sentence. 
The anchor-based context-aware query $S^1$ should be more representative than the query $S^2$ comprised of terms sampled from the destination page.
We train the model to predict the pair-wise preference of the two queries $S^1$ and $S^2$.

Specifically, inspired by the strong ability of BERT~\cite{DBLP:conf/naacl/DevlinCLT19} to aggregate context and model sequences, we firstly use BERT to calculate the contextual word representations of the sentence $S$. Specifically, for a sentence $S=(w_1,w_2,\cdots,w_n)$, we get its contextual representation $H=(h_1,h_2,\cdots,h_n)$, where $h_t$ denotes a $d$-dimension hidden vector of the $t$-th sentence token. Assume that for anchor text $a$ in sentence $S$, the corresponding hidden vector $a$ is $h_a$. We calculate the self-attention weight $\alpha^t$ of each word $w_t$ based on the anchor text $a$ as the average weights across $D$ heads:
\begin{align}
    \alpha^t &= \frac{1}{D}\sum^{D}_{i=1}\alpha_i^t =  \frac{1}{D}\sum^{D}_{i=1}{\rm softmax}(\frac{W^{Q}_{i}h_{a} \cdot W^{K}_{i}h_{t}}{\sqrt{d/D}}) ,
\end{align}
where $\alpha_i^t$ is the attention weight on the $i$-th head.
Typically, a term may appear multiple times within the same sentence. Thus, we add up the attention weights of the same tokens over different positions in the sentence $S$. Specifically, for each distinct term $w_k$ in the vocabulary $V=\{w_k\}_{k=1}^{K}$, we calculate the final weight of distinct token $w_k$ as:
\begin{align}
    \beta_{w_k} &= \sum_{w_t=w_k}\alpha^t.\label{merge_rep}
\end{align}
Finally, we normalize the distinct weights of all terms in the vocabulary to obtain a distribution $p(w_k)$ across the terms as:
\begin{align}
    p(w_k) &= \frac{{\rm exp}(\beta_{w_k})}{\sum_{w_k \in V}{\rm exp}(\beta_{w_k})}. \label{softmax}
\end{align}
The term distribution can measure the contextual similarity between the word $w_k$ and anchor text $a$. Thus, we use this distribution for sampling $l$ words from the sentence $s$ to form the query $S^1$. Based on this distribution, the words relevant to the anchor text can be sampled with a higher probability. Thus, we can build a query $S^1$ based on the reliable signals of the anchor texts. Following ~\cite{10.1145/3437963.3441777,DBLP:conf/sigir/AzzopardiRB07, DBLP:conf/sigir/MaGZFLC21}, the size $l$ of the pseudo query is calculated through a Poisson distribution as:
\begin{align}
   P(x) &= \frac{\lambda^x e^{-\lambda}}{x}, x=1,2,3,\cdots.
\end{align}
Finally, we collect the $l$ sampled words and the anchor text $a$ together, and construct a word set $S^1$ of length $l$+1. Since $S^1$ is formed from the anchor text $a$ and its contextual words, there could be high relevance between this word set and destination page of $a$. 

For constructing the pair-wise loss, we also need to construct the negative query. Since proper hard negatives can help to train a better ranking model~\cite{DBLP:journals/corr/abs-2007-00808,DBLP:conf/emnlp/KarpukhinOMLWEC20}, we propose to sample representative words from the destination page $P$ to construct the hard negative pseudo query for the page, rather than sample words from unrelated pages randomly. For selecting representative words to build the negative query, we firstly use BERT to generate the contextual representations of $P=(p_1,p_2,...)$ as $(h^P_1,h^P_2...)$, where $h^P_t$ is the hidden state of the $t$-th term in $P$. Then, we also use the self-attention weights to measure the sampling probability of terms in page $P$. Unlike the phrase for constructing $S^1$, we calculate the self-attention weights of each terms based on the special token [CLS] as: 
\begin{align}
\alpha_{i}^{t} &= \begin{cases}
{\rm softmax}(\frac{W^{Q}_{i}h_{\rm [CLS]} \cdot W^{K}_{i}h^P_t}{\sqrt{d/D}}), & P_t \neq a, \\
0, & P_t=a.
\end{cases}
\end{align}
For the term in anchor text $a$, we set their weight to 0. Thus, the term in anchor text will not be selected into $S^2$, and the relevance signal of anchor text in $S^1$ will indeed enhance the Transformer model. We then perform sum operation for repetitive words following Equation~(\ref{merge_rep}), normalize the term distribution following Equation~(\ref{softmax}), and generate the word set $S^2$ from passage $P$. The word set $S^2$ generated from passage $P$ will be used as the negative query.

Finally, we formulate the objective of the Representative Query Prediction task by a typical pairwise loss, \ie, hinge loss for the pre-training as: 
\begin{align}
    \mathcal{L}_{\rm RQP} &= \max(0, 1 - p(S_1|P) + p(S_2|P)),
\end{align}
where $p(S|P)$ is the matching score between the word set $S$ and the page $P$. We concatenate the word set $S$ and $P$ as a single input sequence and feed into the the Transformer with delimiting tokens [SEP]. Then, we calculate the matching score by applying a MLP function over the classification token's representation as:
\begin{align}
    p(S|P) &= \rm{MLP}(h^{[\rm{CLS}]}), \label{cls_cal}\\
    h^{[\rm{CLS}]} &= {\rm Transformer}([\rm{CLS}]+S+[\rm{SEP}]+P+[\rm{SEP}]).\label{bert_concat}
\end{align}

\subsubsection{Query Disambiguation Modeling~(QDM)}

In real-world applications, the queries issued by users are often short and ambiguous~\cite{DBLP:journals/sigir/SilversteinHMM99,DBLP:conf/wsdm/ZhouDW20,DBLP:conf/sigir/ZhouDWXW21,DBLP:conf/cikm/MaDBW20}, such as the query ``Apple''~(Apple fruit or Apple company?).  
Thus, building an accurate encoding of the input query is difficult, which further leads to the poor quality of these ambiguous queries.
Fortunately, with hyperlinks and anchor texts, we can endow the language representation model with the ability to disambiguate queries in the pre-training stage. We observe that the same anchor texts could link to one or more different pages. 
Under these circumstances, the anchor could be viewed as an ambiguous query, while the context around the anchor text could help to disambiguate the query.
We train the model to predict the true destination page with the semantic information brought by the query context, thus learn disambiguation ability while pre-training. 

Specifically, For each distinct anchor text $a$, we collect all of its occurrence in corpus $\mathcal{C}$ as $C^a = ((a,S_1,P_1),\cdots,(a,P_{|a|},p_{|a|}))$, where $(a,S,P)$ means that the destination page of $a$ is $P$ when $a$ is in sentence $S$. Assume that for one occurrence $(a,S,P)$, following Section~\ref{sec:RQP}, we can build a context-aware word set $S^1$ from sentence $S$ based on anchor text $a$. We treat $S^1$ as the query, page $P$ as the relevant document.
Then we sample a negative page $\tilde{P}$ from the pages $(P_1,\cdots, P_{|a|})$.
We train the model to predict the pair-wise matching preference between the query $S^1$ and the two pages as:
\begin{align}
    \mathcal{L}_{\rm QDM} &= \max(0, 1 - p(S_1|P) + p(S_1|\tilde{P})),
\end{align}
where $p(S|P)$ follows the [CLS] score calculation in Equation~(\ref{cls_cal}).
As illustrated above, the anchor-based contextual word set $S^1$ sampled from the sentence $S$ can provide additional semantic information for the pre-trained model. Thus, even the anchor text has also pointed to the negative sample, the model can be trained to learn the fine-grained relevance based on the context information around the anchor text. By leveraging the context, the model will learn the ability to query disambiguation.

\subsubsection{Representative Document Prediction~(RDP)}

Although most queries presented to search engines vary between one to three terms in length, a gradual increase in the average query length has been observed in recent studies~\cite{DBLP:conf/sigir/KumaranC09,DBLP:conf/sigir/DattaV11,DBLP:conf/sigir/BalasubramanianKC10}. 
Even though these queries could convey more sophisticated information needs of users, they also carry more noises to the search engine. 
A common strategy to deal with the long queries is to let the model distinguish the important terms in the queries, then focus more on these terms to improve retrieval effectiveness~\cite{DBLP:conf/sigir/BenderskyC08,DBLP:conf/wsdm/BenderskyMC10,DBLP:conf/ecir/LeaseAC09}. 
At the pre-training stage of ad-hoc retrieval, if the model can be trained with more samples with long queries, it will get more robust when fine-tuning for long queries. 
Besides, the language model should be trained to predict the most representative document for the long query, since the long query could focus on different views.
Fortunately, the hyperlinks can help to build pre-training samples containing long queries. We observe that there could be more than one anchor text appearing in one sentence. If we treat the sentence as the query, the destination pages could be the relevant documents for the sentence. Besides, if the anchor text is more important in the sentence, its destination page would be more representative for the sentence. Inspired by this, we propose to pre-train the model to predict the relevant document for the sentence containing more than one hyperlinks.

Specifically, for sentence $S = (w_1,w_2,\cdots,w_n)$ and its anchor texts set $A = ((a_1,P_1), (a_2,P_2),\cdots, (a_m,P_m))$, we sample two anchors based on the anchor importance of this sentence. We will treat the sentence $S$ as a long query, and the two destination pages as the documents. 
In this way, the page is deemed as a more representative document if its anchor text is of higher importance. To measure the importance of the anchor text, we use the Transformer encoder to build the context-aware representations of the terms, and calculate the hidden vectors $H=(h_1,h_2,\cdots,h_n)$. Assume that for anchor text $a$, its hidden vector is $h_a$. We calculate the self-attention weight of anchor $a$ based on the classification token [CLS] to measure its importance as:
\begin{align}
    \gamma^a &= \frac{1}{D}\sum^{D}_{i=1}\gamma_i^a = \frac{1}{D}\sum^{D}_{i=1}{\rm softmax}(\frac{W^{Q}_{i}h_{a} \cdot W^{K}_{i}h_{{\rm [CLS]}}}{\sqrt{d/D}}),
\end{align}
where we average the attention weights across $D$ heads. The token [CLS] is an aggregate of the entire sequence representation, and it can represent the comprehensive understanding of the input sequence over all tokens. Thus, the attention weight $\gamma^a$ could measure the contribution of the anchor $a$ to the entire sentence. Then, we merge the repeat anchor texts in one sentence following Equation~(\ref{merge_rep}), normalize the weights to a probability distribution $p(a)$ over all anchor texts following Equation~(\ref{softmax}) as:
\begin{align}
    p(a) = \frac{{\rm exp}(\eta_{a})}{\sum_{a \in A}{\rm exp}(\eta_{a})}, \quad \eta_{a} = \sum_{a_t=a_k}\gamma^a,  \label{sample_prob}
\end{align}
According to the importance likelihood $p(a)$ of anchors, we sample two anchor texts $(a_1,P_1)$ and $(a_2,P_2)$ from the sentence $S$. Suppose that $a_1$ has a higher importance likelihood than $a_2$ according to Equation~(\ref{sample_prob}). We treat the sentence $S$ as the long query, $P_1$ as the more representative page and $P_2$ as the less representative page. We minimize the pair-wise loss $\mathcal{L}_{RDP}$ by:
\begin{align}
    \mathcal{L}_{\rm RDP} &= \max(0, 1 - p(S|P_1) + p(S|P_2)),
\end{align}
where $p(S|P)$ follows the similar calculation in Equation~(\ref{cls_cal}).

\subsubsection{Anchor Co-occurrence Modeling~(ACM)}

Language models try to learn the term semantics by modeling the term co-occurrence relation, including the term co-occurrence in one window~\cite{DBLP:conf/emnlp/PenningtonSM14,DBLP:conf/nips/MikolovSCCD13} and in one sequence~\cite{DBLP:conf/naacl/DevlinCLT19,DBLP:conf/naacl/PetersNIGCLZ18}. As special terms, the anchor texts also share the co-occurrence relations. Besides, since the destination page can help to provide additional information to understand the anchor texts, we can learn more accurate semantics based on the co-occurrence relation by leveraging these destination pages. 
Therefore, we propose the Anchor Co-occurrence Modeling~(ACM) task to model the similarity between the semantics of the anchors in one sentence. 
By pre-training with ACM, the model could obtain similar representations for the co-occurrenced anchor texts in one sentence, which further improves its ability to model semantics.

Suppose that for a sentence $S=(w_1,w_2,\cdots,w_n)$ and its anchor texts set $A = ((a_1,P_1),(a_2,P_2),\cdots,(a_m,P_m))$, the anchors in $A$ all share the co-occurrence characteristics with each other. We sample a pair of anchors randomly as $(a_1,P_1)$ and $(a_2,P_2)$. We then sample some important words from the page $P_1$ to form a word set $S^1$. Then, we let the model learn the semantic matching between $S^1$ and the passage $P^2$, thus incorporating the anchor co-occurrence into the pre-trained model. Specifically, for the destination page $P_1$ of anchor $a_1$, we use a Transformer encoder to build contextual representations and use the attention weight of [CLS] to measure the term importance:
\begin{align}
    \mu_{i}^{t} &= {\rm softmax}(\frac{W^{Q}_{i}h_{\rm [CLS]} \cdot W^{K}_{i}p_{t}}{\sqrt{d/D}}),
\end{align}
where $\mu_{i}^{t}$ is the attention weight of $t$-th term based on [CLS] token on the $i$-th head. Then, we average the term weights across all heads as $\mu^t = \frac{1}{D}\sum^{D}_{i=1}\mu_i^t$.
After merging repetitive terms and normalization following Section~\ref{sec:RQP}, we sample words based on the final word importance probabilities. Then, these sampled words form the word set $S^1$ jointly with the anchor text $a_1$. Since the anchor set $S^1$ reflects the information of anchor text $a_1$, we use $S^1$ and $P_2$ as the query and the relevant document to learn the semantic matching degree, respectively. We sample a page $\tilde{P}$ from the corpus $\mathcal{C}$ as the negative document, then learn the pair-wise loss of Anchor Co-occurrence Modeling task as:
\begin{align}
    \mathcal{L}_{\rm ACM} &= \max(0, 1 - p(S_1|P_2) + p(S_1|\tilde{P})).
\end{align}

\subsubsection{Final Training Objective} \label{sec:final_objective}

Besides the pair-wise loss to measure the relevance between pseudo queries and documents, the pre-trained model also needs to build good contextual representations for them. Following~\cite{10.1145/3437963.3441777,DBLP:conf/naacl/DevlinCLT19}, we also adopt the Masked Language Modeling~(MLM) as one of our objectives. MLM is a fill-in-the-blank task, which firstly masks out some tokens from the input, then trains the model to predict the masked tokens by the rest tokens. Specifically, the MLM loss is defined as:
\begin{align}
    \mathcal{L}_{\rm MLM} &= -\sum_{\hat{x} \in m(X)} \log p(\hat{x} | x_{\backslash m(X)}),
\end{align}
where $X$ denote the input sentence, and $m(X)$ and $x_{\backslash m(X)}$ denotes the masked tokens and the rest tokens from $X$, respectively. 

Finally, we pre-train the Transformer model $\mathcal{M}$ towards the proposed four objectives jointly with the MLM objective as: 
\begin{align}
    \mathcal{L} &= \mathcal{L}_{\rm RQP} + \mathcal{L}_{\rm QDM} + \mathcal{L}_{\rm RDP} + \mathcal{L}_{\rm ACM} + \mathcal{L}_{\rm MLM}.
\end{align}
All parameters are optimized by the loss $\mathcal{L}$, and the whole model is trained in an end-to-end manner.

\subsection{Fine-tuning for Document Ranking} \label{sec:fine-tune}
In the previous pre-training stage, we pre-train the Transformer model $\mathcal{M}$ to learn the IR matching from the raw corpus based on the hyperlinks and anchor texts. We now incorporate $\mathcal{M}$ into the downstream document ranking task to evaluate the effectiveness of our proposed pre-trained method. 

Previous studies have explored utilizing Transformer to measure the sequence pair relevance for ad-hoc document ranking~\cite{DBLP:journals/corr/abs-1910-14424,DBLP:journals/corr/abs-1901-04085,qiao2019understanding}. For the query $q$ and a candidate document $d$, we aim to calculate a ranking score $s(q,d)$ to measure the relevance between them based on the pre-trained Transformer. Therefore, in this stage, we firstly use the same Transformer architecture as the pre-trained model $\mathcal{M}$, and use the parameters and embeddings of $\mathcal{M}$ to initialize the Transformer model. Then, we add special tokens and concatenate the query and the document as $Y=([{\rm CLS}];q;[{\rm SEP}];d;[{\rm SEP}])$, where $[;]$ is the concatenation operation. A $[{\rm SEP}]$ token is added at the tail of query and document, while a $[{\rm CLS}]$ token is added at the sequence head for summary. Finally, We feed the concatenated sequence into Transformer, and use the $[{\rm CLS}]$ representation $z^{\rm [CLS]}$ to calculate the final ranking score as:
\begin{align}
    s(q,d) = {\rm MLP}(z^{\rm [CLS]}), \quad z^{\rm [CLS]} = {\rm Transformer}(Y).
\end{align}
To train the model, we use the cross-entropy loss for optimization:
\begin{align}
    \mathcal{L}_{rank} &= \frac{1}{N} \sum_{i=1}^{N} y_i\log(s(q,d)) + (1-y_i)\log(1-s(q,d)),
\end{align}
where $N$ is the number of samples in the training set.

\begin{table}[t!]  
\small
\caption{The data statistics of four pre-training tasks.}  
\label{table:data}
\begin{tabular}{lrrrr}  
\toprule
Tasks & \#tokens  & \#pairs & avg. \#query tokens &  \#doc tokens \\  
\midrule
  RQP & 3.33B & 17.71M & 2.64  & 90.48  \\
  QDM & 0.22B  & 1.35M & 2.80  & 81.17  \\
  RDP & 1.56B &  6.49M & 34.90  & 85.86   \\
  ACM & 1.97B  & 12.64M & 4.55  & 73.64  \\
\bottomrule  
\end{tabular}  
\end{table}

\section{Experiments}\label{sec:exp}
\subsection{Datasets and Evaluation Metrics}\label{sec:datasets}

\subsubsection{Pre-training Corpus}
We use English Wikipedia (2021/01/01)\footnote{\url{https://dumps.wikimedia.org/enwiki/}} as the pre-training corpus, since they are publicly available and have a large-scale collection of documents with hyperlinks for supporting pre-training. Following~\cite{DBLP:conf/coling/SunSQGHHZ20,10.1145/3437963.3441777}, we use the public WikiExtractor\footnote{\url{https://github.com/attardi/wikiextractor}} to process the download Wikipedia dump while preserving the hyperlinks. After removing the articles whose length is less than 100 words for data cleaning, it comprises 15,492,885 articles. The data for our proposed four tasks are generated from these articles, and the statistics are reported in Table~\ref{table:data}.
We pre-train the model on one combined set of query-document pairs, where each pair is uniformly sampled from the four pre-training tasks.

\subsubsection{Fine-tuning Datasets}

To prove the effectiveness of the proposed pre-training methods, we conduct fine-tuning experiments on two representative ad-hoc retrieval datasets. 
\begin{itemize}[leftmargin=*]
\item \textbf{MS MARCO Document Ranking~(MS MARCO)}\footnote{\url{https://github.com/microsoft/MSMARCO-Document-Ranking}}~\cite{msmarco_dataset}: It is a large-scale benchmark dataset for document retrieval task. It consists of 3.2 million documents with 367 thousand training queries, 5 thousand development queries, and 5 thousand test queries. The relevance is measured in 0/1.
\item \textbf{TREC 2019 Deep Learning Track~(TREC DL)}\footnote{\url{https://microsoft.github.io/msmarco/TREC-Deep-Learning-2019.html}}~\cite{trecdl_dataset}: It replaces the test queries in MS MARCO with a novel set with more comprehensive notations. Its test set consists of 43 queries, and the relevance is scored in 0/1/2/3.
\end{itemize}

\subsubsection{Evaluation Metrics} Following the official instructions, we use MRR@100 and nDCG@10 to measure and evaluate the top-ranking performance. Besides, we also calculate MRR@10 and nDCG@100 for MS MARCO and TREC DL, respectively.

\begin{table*}[t!]
    \centering
    \small
    \caption{Evaluation results of all models on two large-scale datasets. ``$\dagger$'' denotes the result is significantly worse than our method HARP in t-test with $p \textless 0.05$ level. The best results are in  \textbf{bold} and the second best results are \underline{underlined}.}
    \label{table_a}
    \begin{tabular}{clcccccccc}
    \toprule
     \multirow{3}[3]{*}{Model Type} & \multirow{3}[3]{*}{Model Name} & \multicolumn{4}{c}{MS MARCO} & \multicolumn{4}{c}{TREC DL} \\
         \cmidrule(lr){3-6} \cmidrule(lr){7-10}
        & & \multicolumn{2}{c}{ANCE Top100} & \multicolumn{2}{c}{Official Top100} & \multicolumn{2}{c}{ANCE Top100} & \multicolumn{2}{c}{Official Top100} \\
        \cmidrule(lr){3-4} \cmidrule(lr){5-6} \cmidrule(lr){7-8} \cmidrule(lr){9-10}
        & & MRR@100 & MRR@10 & MRR@100 & MRR@10 & nDCG@100 & nDCG@10 & nDCG@100 & nDCG@10 \\
        
        \midrule
        \multirow{2}{*}{\minitab[c]{Traditional \\ IR Models}} &  QL & $.2457^\dagger$ & $.2295^\dagger$ & $.2103^\dagger$ & $.1977^\dagger$ & $.4644^\dagger$ & $.5370^\dagger$ & $.4694^\dagger$ & $.4354^\dagger$ \\
        &  BM25 & $.2538^\dagger$ & $.2383^\dagger$ & $.2379^\dagger$ & $.2260^\dagger$ & $.4692^\dagger$ & $.5411^\dagger$ & $.4819^\dagger$ & $.4681^\dagger$ \\
        
        \midrule
         \multirow{4}{*}{\minitab[c]{Neural \\ IR Models}} &  DRMM & $.1146^\dagger$ & $.0943^\dagger$ & $.1211^\dagger$ & $.1047^\dagger$ & $.3812^\dagger$ & $.3085^\dagger$ & $.4099^\dagger$ & $.3000^\dagger$ \\
        &  DUET & $.2287^\dagger$ & $.2102^\dagger$ & $.1445^\dagger$ & $.1278^\dagger$ & $.3912^\dagger$ & $.3595^\dagger$ & $.4213^\dagger$ & $.3432^\dagger$ \\
        &  KNRM & $.2816^\dagger$ & $.2740^\dagger$ & $.2128^\dagger$ & $.1992^\dagger$ & $.4671^\dagger$ & $.5491^\dagger$ & $.4727^\dagger$ & $.4391^\dagger$ \\
        &  Conv-KNRM & $.3182^\dagger$ & $.3054^\dagger$ & $.2850^\dagger$ & $.2744^\dagger$ & $.4876^\dagger$ & $.5990^\dagger$ & $.5221^\dagger$ & $.5899^\dagger$ \\
        
        \midrule
        \multirow{6}{*}{\minitab[c]{Pretrained \\ IR Models}} &  BERT & $.4184^\dagger$ & $.4091^\dagger$  & $.3830^\dagger$ & $.3770^\dagger$ & $.4900$ & $.6084^\dagger$ & $.5289^\dagger$ & $.6358^\dagger$ \\
        &  $\rm{Transformer}_{ICT}$ & $.4194^\dagger$ & $.4101^\dagger$ & $.3828^\dagger$ & $.3767^\dagger$ & $\underline{.4906}$ & $.6053^\dagger$ & $\underline{.5300}$ & $\underline{.6386}^\dagger$ \\
        &  $\rm{Transformer}_{WLP}$ & $.3998^\dagger$ & $.3900^\dagger$ & $.3698^\dagger$ & $.3635^\dagger$ & $.4891^\dagger$ & $.6143$ & $.5245^\dagger$ & $.6276^\dagger$\\
        &  $\rm{PROP}_{Wiki}$ & $.4188^\dagger$ & $.4092^\dagger$ & $.3818^\dagger$ & $.3759^\dagger$ & $.4882^\dagger$ & $.6050^\dagger$ & $.5251^\dagger$ & $.6224^\dagger$ \\
        &  $\rm{PROP}_{MARCO}$ & $\underline{.4201}^\dagger$ & $\underline{.4111}^\dagger$ & $\underline{.3856}^\dagger$ & $\underline{.3800}^\dagger$ & $.4894$ & $\underline{.6166}$ & $.5242^\dagger$ & $.6208^\dagger$ \\
        &  HARP (ours) & $\textbf{.4472}$ & $\textbf{.4393}$ & $\textbf{.4012}$ & $\textbf{.3961}$ & $\textbf{.4949}$ & $\textbf{.6202}$ & $\textbf{.5337}$ & $\textbf{.6562}$ \\
    \bottomrule
    \end{tabular}
\end{table*}

\begin{table}[t!]
    \centering
    \small
    \caption{Document Ranking Performance measured on MS MARCO leaderboard. As the leaderboard only reports aggregated metrics, we cannot report statistical significance.}
    \label{table_leaderboard}
    \begin{tabular}{lcc}
    \toprule
        Method & Dev MRR@100 & Eval MRR@100 \\
    \midrule
        PROP (ensemble v0.1) & .4551 & .4010   \\
        BERT-m1 (ensemble) & .4633 & .4075\\
        LCE Loss (ensemble) & .4641 & .4054 \\
    \midrule
        HARP (single)  & .4472 & .3895 \\
        HARP (ensemble)  & \textbf{.4711} & \textbf{.4159} \\
    \bottomrule
    \end{tabular}
\end{table}

\begin{table}[t!]
    \centering
    \small
    \caption{Evaluation results of ablation models. ``$\dagger$'' denotes the result is significantly worse than our method HARP in t-test with $p \textless 0.05$ level. The best results are in \textbf{bold}.}
    \label{table_ablation}
    \begin{tabular}{lcccc}
    \toprule
        \multirow{2}[2]{*}{Model Name} & \multicolumn{2}{c}{ANCE Top100} & \multicolumn{2}{c}{Official Top100} \\
        \cmidrule(lr){2-3} \cmidrule(lr){4-5} 
        & MRR@100 & MRR@10 & MRR@100 & MRR@10 \\
        \midrule
        BERT & $.4184^\dagger$ & $.4091^\dagger$  & $.3830^\dagger$ & $.3770^\dagger$ \\
        \midrule
        \quad  \textit{w/o}  MLM & $.4435$ & $.4357$ & $.3947^\dagger$ & $.3914^\dagger$   \\
        \quad  \textit{w/o}  RQP & $.4361^\dagger$ & $.4278^\dagger$ & $.3918^\dagger$ & $.3865^\dagger$   \\
        \quad  \textit{w/o}  QDM & $.4423^\dagger$ & $.4340^\dagger$ & $.3931^\dagger$ & $.3879^\dagger$   \\
        \quad  \textit{w/o}  RDP & $.4387^\dagger$ & $.4305^\dagger$ & $.3917^\dagger$ & $.3867^\dagger$   \\
        \quad  \textit{w/o}  ACM & $.4424^\dagger$ & $.4341$ & $.3934^\dagger$ & $.3882^\dagger$  \\
        \midrule
        HARP (ours) &  $\textbf{.4472}$ & $\textbf{.4393}$ & $\textbf{.4012}$ & $\textbf{.3961}$\\
    \bottomrule
    \end{tabular}
\end{table}

\subsection{Baselines}\label{sec:baselines}

We evaluate the performance of our approach by comparing it with three groups of highly related and strong baseline methods:

(1) \textit{Traditional IR models}. 
\textbf{QL}~\cite{DBLP:journals/sigir/ZhaiL17} is one of the best performing models which measure the query likelihood of query with Dirichlet prior smoothing.
\textbf{BM25}~\cite{DBLP:conf/sigir/RobertsonW94} is another famous and effective retrieval method based on the probability retrieval model.

(2) \textit{Neural IR models}. 
\textbf{DRMM}~\cite{DBLP:conf/cikm/GuoFAC16} is a deep relevance matching model which performs histogram pooling on the transition matrix and uses the binned soft-TF as the input to a ranking neural network.
\textbf{DUET}~\cite{DBLP:conf/www/Mitra0C17} propose to use two separate networks to match queries and documents with local and learned distributed representations, respectively. The two networks are jointly trained as part of a single neural network.
\textbf{KNRM}~\cite{DBLP:conf/sigir/XiongDCLP17} is a neural ranking model which extracts the
features of interaction between query and document terms. The kernel-pooling is used to provide soft match signals for ranking.
\textbf{Conv-KNRM}~\cite{DBLP:conf/sigir/XiongDCLP17} is an upgrade of the KNRM model. It adds a convolutional layer for modeling n-gram soft matches and fuse the contextual information of surrounding words for matching.

(3) \textit{Pre-trained Models}. 
\textbf{BERT}~\cite{DBLP:conf/naacl/DevlinCLT19} is the multi-layer bi-directional Transformer pre-trained with Masked Language Modeling and Next Sentence Prediction tasks. 
$\bm{{\rm Transformer}_{\rm ICT}}$~\cite{DBLP:conf/iclr/ChangYCYK20} is the BERT model retrained with the Inverse Cloze Task~(ICT) and MLM. It is specifically
designed for passage retrieval in QA scenarios, which teaches the
model to predict the removed sentence given a context text.
$\bm{{\rm Transformer}_{\rm WLP}}$~\cite{DBLP:conf/iclr/ChangYCYK20} is the BERT model retrained with the Wiki Link Prediction~(WLP) and MLM. It is designed for capturing inter-page semantic relations.
\textbf{PROP}~\cite{10.1145/3437963.3441777} is the state-of-the-art pre-trained model tailored for ad-hoc retrieval. It uses the Representative Words Prediction task for learning the matching between the sampled word sets. We experiment with both of the released models pre-trained on Wikipedia and MS MARCO corpus, \ie, $\rm{PROP}_{Wiki}$ and $\rm{PROP}_{MARCO}$, respectively.



\subsection{Implementation Details}\label{sec:implement}

\subsubsection{Model Architecture}

For our methods HARP, we use the same Transformer encoder architecture as $\rm{BERT}_{base}$ in BERT~\cite{DBLP:conf/naacl/DevlinCLT19}. The hidden size is 768, and the number of self-attention heads is 12.  For a fair comparison, all of the pre-trained baseline models use the same architecture as our model. We use the HuggingFace’s Transformers for the model implementation~\cite{wolf2020huggingfaces}. 

\subsubsection{Pre-training Settings}
For the construction of the pseudo queries, we set the expectation of interval $\lambda$ as 3, and remove the stopwords using the INQUERY stopwords list following~\cite{10.1145/3437963.3441777}. 
We use the first section to denote the destination page because it is usually the description or summary of a long document~\cite{DBLP:conf/iclr/ChangYCYK20,DBLP:journals/corr/abs-1910-14424,DBLP:conf/sigir/DaiC19}.  For the MLM objective, we follow the settings in BERT, where we randomly select 15\% words for prediction, and the selected tokens are (1) the [MASK] token 80\% of the time, (2) a random token 10\% of the time, and (3) the unchanged token 10\% of the time. We use the Adam optimizer with a learning rate of 1e-4 for 10 epochs, where the batch size is set as 128. For the large cost of training from scratch, we use $\rm{BERT}_{base}$ to initialize our method and baseline models. 

\subsubsection{Fine-tuning Settings}

In the fine-tuning stage, the learned parameters in the pre-training stage are used to initialize the embedding and self-attention layers of our model. 
Following the previous works, we only test the performance of our model on the re-ranking stage~\cite{10.1145/3437963.3441777,DBLP:journals/corr/abs-1901-04085}. To test the performance of our models with different quality of the candidate document set, we re-rank the document from the two candidate sets, \ie, ANCE Top100 and Official Top100. ANCE Top100 is retrieved based on the ANCE model proposed by ~\citet{DBLP:journals/corr/abs-2007-00808}, and Official Top100 is released by the official MS MARCO and TREC teams.
While fine-tuning, we concatenate the title, URL and body of one document as the document content. The batch size is set as 128, and the maximum length of the input sequence is 512. We fine-tune for 2 epochs, with a 1e-5 learning rate and a warmup portion 0.1. 
Our code is available online\footnote{\url{https://github.com/zhengyima/anchors}}.

\subsection{Experimental Results}\label{sec:overall}
Since the MS MARCO leaderboard limits the frequency of submissions, we evaluate our method and baseline methods on MS MARCO's development set. For TREC DL, we evaluate the test set of 43 queries. The overall performance on the two datasets is reported in Table~\ref{table_a}. 
We can observe that:

(1) Among all models, \textbf{HARP achieves the best results in terms of all evaluation metrics}. HARP improves performance with a large margin over two strongest baselines $\rm{PROP}_{Wiki}$ and $\rm{PROP}_{MARCO}$, which also design objectives tailored for IR. Concretely, HARP significantly outperforms $\rm{PROP}_{MARCO}$ by 6.4\% in MRR@100 on MS MARCO ANCE Top100. On TREC DL ANCE Top100 in terms of nDCG@100, HARP outperforms $\rm{PROP}_{MARCO}$ by 1.1\%. The reason for the improvement reduction on the TREC DL set is that it uses binary notations in the training set but a multi-label notation in the test set, which leads to a gap and difficulty increase. Besides, HARP outperforms the best baselines for both the ANCE Top100 set and the Official Top100 set. These results demonstrate that HARP can capture better matching under different quality of the candidate list, while not being limited by the lower candidate quality or confused by the harder negatives. All these results prove that introducing hyperlinks into pre-training can improve the ranking quality of the pre-trained language model.

(2) \textbf{All pre-trained methods outperform methods without pre-training}, indicating that pre-training and fine-tuning are helpful for improving the relevance measuring of models for downstream ad-hoc retrieval. Traditional IR models QL and BM25 are strong baselines on the two datasets, but loses the ability to model semantic relevance. Neural IR models use distributed representations to denote the query and document, then apply deep neural networks to measure the IR relevance. Thus, the neural method Conv-KNRM significantly outperforms the traditional methods. The pre-trained methods have dramatic improvements over other methods. This indicates that pre-training on a large corpus and then fine-tuning on downstream tasks is better than training a neural deep ranking model from scratch. 

(3) Among all pre-trained methods, \textbf{the ones designing objectives tailored for IR perform better}.
${\rm Transformer}_{\rm ICT}$ show better performance than BERT, confirming that using a pre-trained task related to retrieval is helpful for downstream tasks. However, $\bm{{\rm Transformer}_{\rm WLP}}$ performs worse than BERT and $\bm{{\rm Transformer}_{\rm ICT}}$. One possible reason is that the queries straightly generated from WLP are noisy since there could be many links in the passage that contribute little to the passage semantics. 
$\rm{PROP}_{Wiki}$ and $\rm{PROP}_{MARCO}$ are the state-of-the-art baselines, which design Representative Words Prediction task tailored for IR. 
Different from the existing objectives, we design four pre-training tasks based on hyperlinks and anchor texts, which bring more accurate and reliable supervised signals. 
Hence, HARP achieves significant improvements compared with the existing pre-trained methods.

Besides, to further prove the effectiveness of HARP, we also report some leaderboard results of MS MARCO on eval set in Table~\ref{table_leaderboard}. We select some representative methods from the leaderboard as the baselines~\cite{10.1145/3437963.3441777,DBLP:conf/ecir/GaoDC21,boytsov2021exploring}. Following other recent leaderboard submissions, we further incorporate model ensemble. Our ensemble entry uses an trained ensemble of using BERT, RoBERTa~\cite{DBLP:journals/corr/abs-1907-11692} and ELECTRA~\cite{DBLP:conf/iclr/ClarkLLM20} to fine-tune the downstream task. 
The leaderboard results confirm the effectiveness of our proposed HARP model.


\subsection{Further Analysis}

We further analyze the influence of different pre-training tasks we proposed~( Section~\ref{sec:ablation}), and the performance under different scales of fine-tuning data~(Section~\ref{sec:finetune-step}).

\subsubsection{Ablation Study}\label{sec:ablation}

Our proposed pre-training approach HARP designs four pre-training objectives to leverage hyperlinks and anchor texts tailored for IR. We remove one of them once a time to analyze its contribution. Note that when none of the pre-training tasks are used, our model degenerates to using BERT for fine-tuning directly. Thus, we also provide the result of BERT for comparison.  We report the MRR@100 and MRR@10 on MS MARCO dataset.

From the results in Table~\ref{table_ablation}, we can observe that removing any pre-training task would lead to a performance decrease. It indicates that all the pre-training tasks are useful to improve the ranking performance. Specifically, removing RQP causes the most decline in all metrics, which confirms that the correlations and supervised signals brought by hyperlinks can improve the ranking ability of our model in the pre-training phase. The significant performance degradation caused by removing RDP shows that pre-training with long queries contributes to further enhancement of ranking relevance modeling. The influence of removing QDM and ACM is relatively smaller. It proves that considering ambiguous query and modeling the anchor co-occurrence are effective but limited, since the pre-training pairs of QDM are less than other tasks, and the queries sampled from the neighboring anchors in ACM are noisier than the anchors. Removing MLM shows the slightest performance decrease, which indicates that good representations obtained by MLM may not be sufficient for ad-hoc retrieval tasks. It is clearly seen that all model variants perform better than BERT, which is not pre-trained by the IR-oriented objectives.

\begin{figure}
\centering
\begin{subfigure}{0.23\textwidth}
\includegraphics[width=0.95\textwidth]{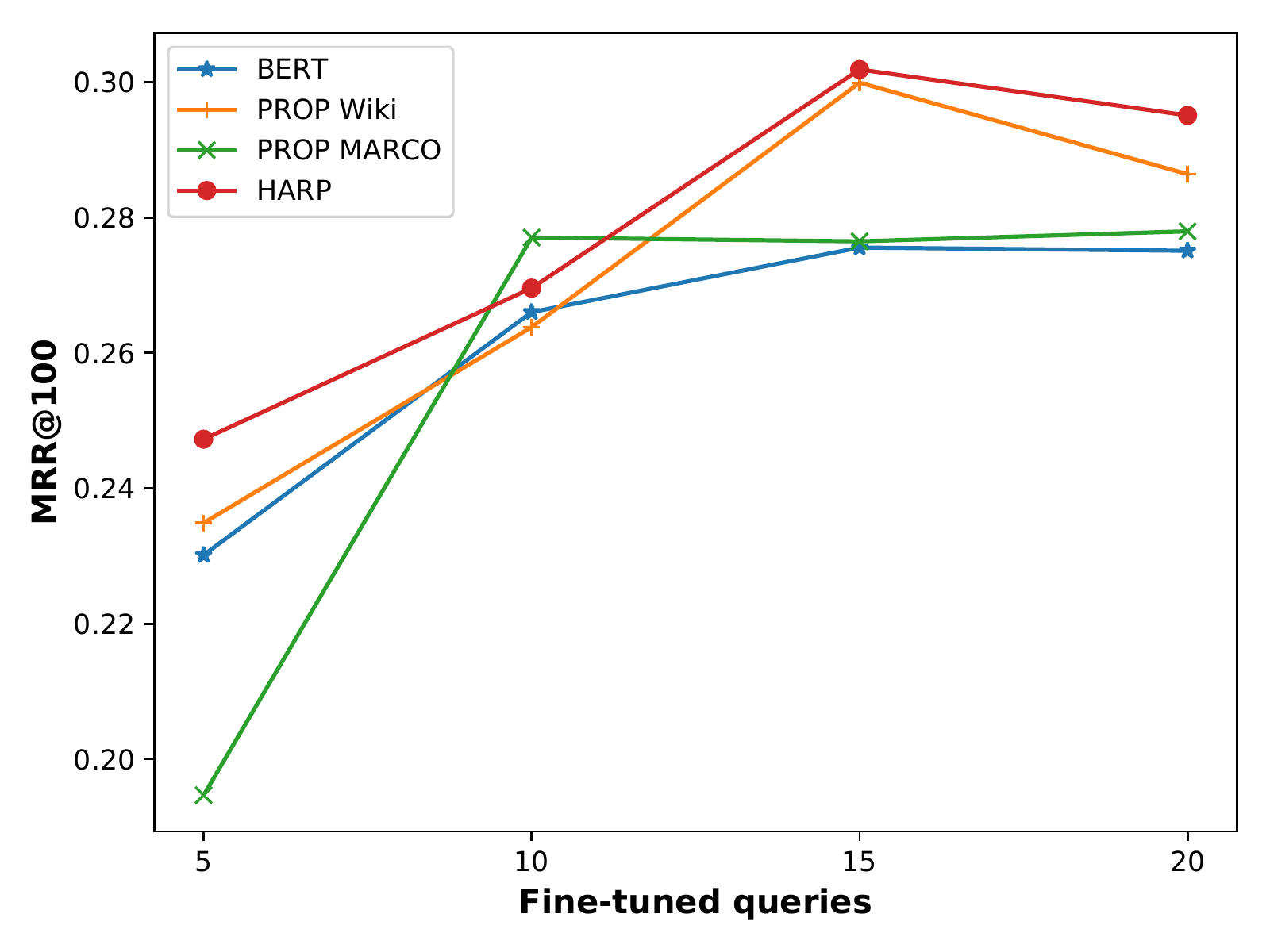}
\caption{Few-shot queries}
\label{fig:fewshot_queries}
\end{subfigure}%
\begin{subfigure}{0.23\textwidth}
\includegraphics[width=0.95\textwidth]{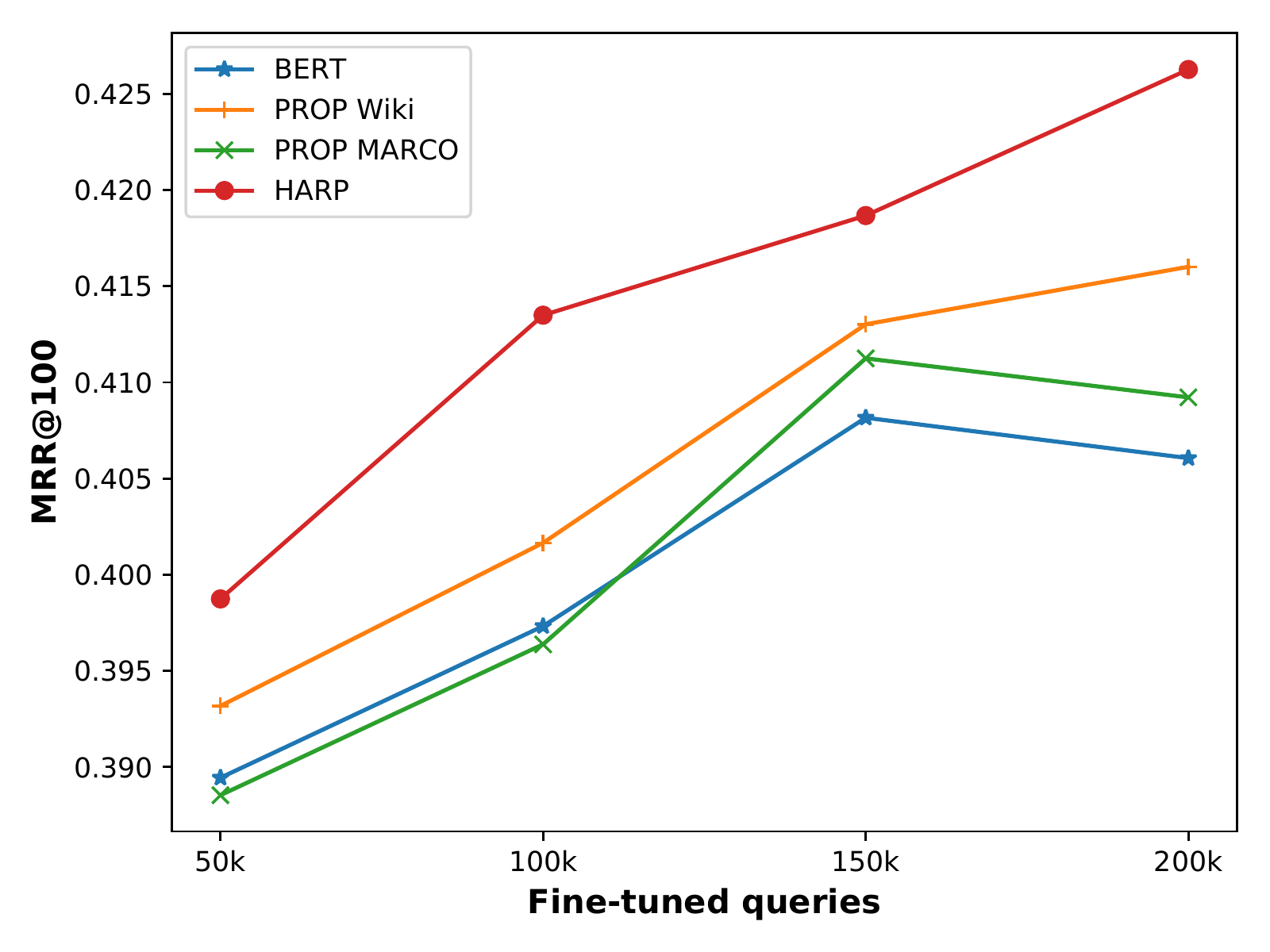}
\caption{large-scale queries}
\label{fig:largescale_queries}
\end{subfigure}%
\caption{Performance on different scales of fine-tune data}\label{figure:difference_scale_quereis}
\end{figure}

\subsubsection{Low-Resource and Large-scale Settings}\label{sec:finetune-step}

Neural ranking models require a considerable amount of training data to learn the representations and matching features. Thus, they are likely to suffer from the low-resource settings in real-world applications, since collecting relevant labels for large-scale queries and documents is costly and time-consuming. This problem can be alleviated by our proposed method, because the pre-training tasks based on hyperlinks and anchor texts can better measure the matching features and resemble the downstream retrieval tasks. To prove that, we simulate the sparsity scenarios by using different scales of queries. For low-resource settings, we randomly pick 5/10/15/20 queries and fine-tune our model. Besides, we also pick 50k/100k/150k/200k queries to evaluate the performance on different large-scale queries. We report MRR@100 to evaluate the performance. We find:

(1) As shown in Figure~\ref{fig:fewshot_queries}, under few-shot settings, HARP can achieve better results compared to other models, showing the scalability for a small number of supervised data. This is consistent with our speculation as tailoring pre-training objectives for IR can provide a solid basis for fine-tuning, which alleviates the influence of data sparsity problem for ranking to some extent.

(2) As shown in Figure~\ref{fig:largescale_queries}, under large-scale settings, HARP is consistently better than baselines in all cases. This further proves the effectiveness of our proposed methods to introduce hyperlinks and anchor texts for designing pre-training objectives for IR. 

(3) When there are large-scale queries, HARP stably performs better when more queries can be used for training. This implies that HARP is able to make better use of fine-tuning data based on the better understandings of IR learned from the pre-training stage.

\section{Conclusion}\label{sec:conclusion}

In this work, we propose a novel pre-training framework HARP tailored for ad-hoc retrieval. Different from existing pre-training objectives tailored for IR, we propose to leverage the supervised signals brought by hyperlinks and anchor texts. We devise four pre-training tasks based on hyperlinks, and capture the anchor-document correlations in different views. We pre-train the Transformer model to predict the pair-wise loss functions built by the four pre-training tasks, jointly with the MLM objective. To evaluate the performance of the pre-trained model, we fine-tune the model on the downstream document ranking tasks. Experimental results on two large-scale representative and open-accessed datasets confirm the effectiveness of our model on document ranking.

\begin{acks}
Zhicheng Dou is the corresponding author. This work was supported by National Natural Science Foundation of China No. 61872370 and No. 61832017,  Beijing Outstanding Young Scientist Program NO. BJJWZYJH012019100020098, Shandong Provincial Natural Science Foundation under Grant ZR2019ZD06, and Intelligent Social Governance Platform, Major Innovation \&  Planning Interdisciplinary Platform for the "Double-First Class" Initiative, Renmin University of China. We also wish to acknowledge the support provided and contribution made by Public Policy and Decision-making Research Lab of RUC.
We are grateful to Yutao Zhu for suggestions on this paper. Finally, comments from the four anonymous referees are invaluable for us to prepare the final version.
\end{acks}

\newpage
\balance

\end{document}